\documentclass[doublecol]{epl2} 

\usepackage{graphicx}
\usepackage{amsmath}
\usepackage{amssymb}
\usepackage{wasysym}
\usepackage{marvosym}
\usepackage{xcolor}
\usepackage{relsize}
\usepackage{bm} 
\usepackage[font=small,skip=0pt]{caption}

\definecolor{ao(english)}{rgb}{0.0, 0.5, 0.0}
\definecolor{verdigris}{rgb}{0.26, 0.7, 0.68}
\definecolor{amber}{rgb}{1.0, 0.75,0.0} 

\newcommand{\mytilde}{\raise.17ex\hbox{$\scriptstyle\mathtt{\sim}$}}

\title{Multiplexing induced explosive synchronization in Kuramoto oscillators with inertia}
\shorttitle{Multiplexing induced explosive synchronization} 

\author{Ajay Deep Kachhvah \and Sarika Jalan}

\institute{Complex Systems Lab, Physics Discipline, Indian Institute of Technology Indore, Simrol, Indore-453552, India}

\abstract{
Explosive synchronization (ES) of coupled oscillators on networks is shown to be originated from existence of correlation between natural frequencies of oscillators and degrees of corresponding nodes. Here, we demonstrate that ES is a generic feature of multiplex network of second-order Kuramoto oscillators and can exist in absence of a frequency-degree correlation. A monoplex network of second-order Kuramoto oscillators bearing homogeneous (heterogeneous) degree-distribution is known to display the first-order (second-order) transition to synchronization. We report that multiplexing of two such networks having homogeneous degree-distribution support the first-order transition in both the layers thereby facilitating ES. More interesting is the multiplexing of a layer bearing heterogeneous degree-distribution with another layer bearing homogeneous degree-distribution, which induces a first-order (ES) transition in the heterogeneous layer which was incapable of showing the same in the isolation. Further, we report that such induced ES transition in the heterogeneous layer of multiplex networks can be controlled by varying inter and intra-layer coupling strengths. Our findings emphasize on importance of multiplexing or impact of one layer on dynamical evolution of other layers of systems having inherent multiplex or multilevel architecture.}

\begin{document}

\maketitle

\section{Introduction}

Networks of coupled oscillators provide a useful paradigm for understanding diverse processes such as epidemic spreading \cite{barrat}, random walks \cite{noh}, traffic congestion \cite{barrat, bocca} and synchronization \cite{pecora, arenas}.\
One of the important factors influencing synchronization of coupled oscillators is topology of the underlying network connecting them \cite{arenas,peron,moreno,lee,arenas2}. It is well known that Kuramoto oscillators display second order transition to the synchronous state \cite{arenas}. In recent years, a new phenomenon called explosive (discontinuous) synchronization has attracted attention of many researchers. Explosive synchronization (ES) had been considered to be an outcome of microscopic correlation between the network topology and natural frequencies of the Kuramoto phase oscillators \cite{gardenes,gardenes2,leyva,peron2,peron3}. 
However, Tanaka \etal \cite{tanaka,tanaka2}, in their seminal work pointed out that a network of second order Kuramoto oscillators i.e.,\ Kuramoto oscillators with an inertia term, is capable of displaying a discontinuous (first-order) transition to synchronous state inherently in absence of any correlation between natural frequencies and network degrees of oscillators.\ Furthermore, recently Peng \etal \cite{peng, peng2} demonstrated that scale-free networks of second-order Kuramoto oscillators with frequency-degree correlation exhibit cluster explosive synchronization.\ Additionally, similar to the first-order Kuramoto oscillators \cite{gardenes}, the discontinuous transition can emerge in a strongly assortative scale-free network of the second-order Kuramoto oscillators in presence of a positive correlation between their degrees and the natural frequencies\cite{peron4}.
 
Further, to fully investigate properties of a diverge range of real-world networks such as transportation \cite{gallotti}, social \cite{szell}, brain \cite{bullmore} or infrastructure \cite{kurant}, a multilayer or multiplex framework \cite{bocca2,leyva2,camellia,shinde,aradhana2} approach is becoming necessary.\ Recently it has been demonstrated that adaptive and multilayer networks of first-order Kuramoto oscillators display ES in absence of a degree-frequency correlation \cite{zhang}.\ In this paper, we study phase transition in a two-layered multiplex network of second-order Kuramoto oscillators. It is shown that the second-order Kuramoto oscillators on a monoplex network bearing homogeneous degree-distribution such as globally connected, random and small-world exhibit the first-order synchronization transition. We show that when a network bearing homogeneous degree-distribution is multiplexed with another network having the same type of degree-distribution, both networks display first-order (explosive) synchronization transition. More interesting is the behavior depicted by the networks having heterogeneous degree-distribution. A monoplex network of second-order Kuramoto oscillators having heterogeneous degree-distribution such as scale-free network displays a second-order phase transition, however, when multiplexed with a network having homogeneous degree-distribution, it starts displaying a first-oder (explosive) synchronization transition. Hence, our investigation reveals that the first-order transition in a network can be induced via multiplexing it with another layer which already exhibits second-order transition.
Further, for the scale-free networks, such induced first-order transition upon multiplexing is weak and can be enhanced by controlling multiplexing properties such as inter and intra-layer coupling strengths.\ Impact of the inter-layer coupling strength on synchronizability in multiplex network is also reported by Dwivedi \etal \cite{sanjiv}.\ We demonstrate that an increase in the inter-layer coupling leads to a gradual strengthening of the first-order (ES) nature of the transition. On the contrary, increasing the intra-layer coupling strength in the scale-free layer yields already weak first-order transition further weaker and eventually reverting it to the second-order.

\section{Theoretical Framework}
For a monoplex network of $N$ second-order Kuramoto phase oscillators, the evolution of each oscillator is ruled by
\begin{align}
m\ddot\theta_{i}(t) + \dot\theta_{i}(t) = \omega_{i} + \frac{\lambda}{\langle k\rangle} \sum_{j=1}^{N} a_{ij} \sin[\theta_{j}(t)-\theta_{i}(t)],
\end{align}
where $i{=}1,...,N$, $\omega_{i}$ ($\theta_{i}(t)$) is the natural frequency (the instantaneous phase) of $i$th oscillator, parameter $m$ is mass, $\lambda$ is homogeneous coupling strength and $\langle k\rangle$ is the average degree of the network.\ Elements of the network's adjacency matrix are indicated by $a_{ij}$, $a_{ij}{=}1$ when the nodes $i$ and $j$ are connected and $a_{ij}{=}0$, otherwise.\ The extent of synchronization in a network can be measured by a global order parameter $R$ defined as $Re^{\iota\psi}{=}(1/N)\sum_{j=1}^{N}e^{\iota\theta_j}$, where $\psi$ is network's average phase, and $0{\leq} R{\leq}1$. The order parameter $R$ takes value $0$ for an incoherent state and $1$ for a phase synchronized state. Upon progressively increasing the coupling strength $\lambda$, the coupled oscillators undergo a change in phase-state from the incoherent to the synchronous one at a critical coupling strength. This process is referred as phase transition to synchronization.

\begin{figure}[t]
	\begin{center}
		\includegraphics[height=3.5cm,width=8cm]{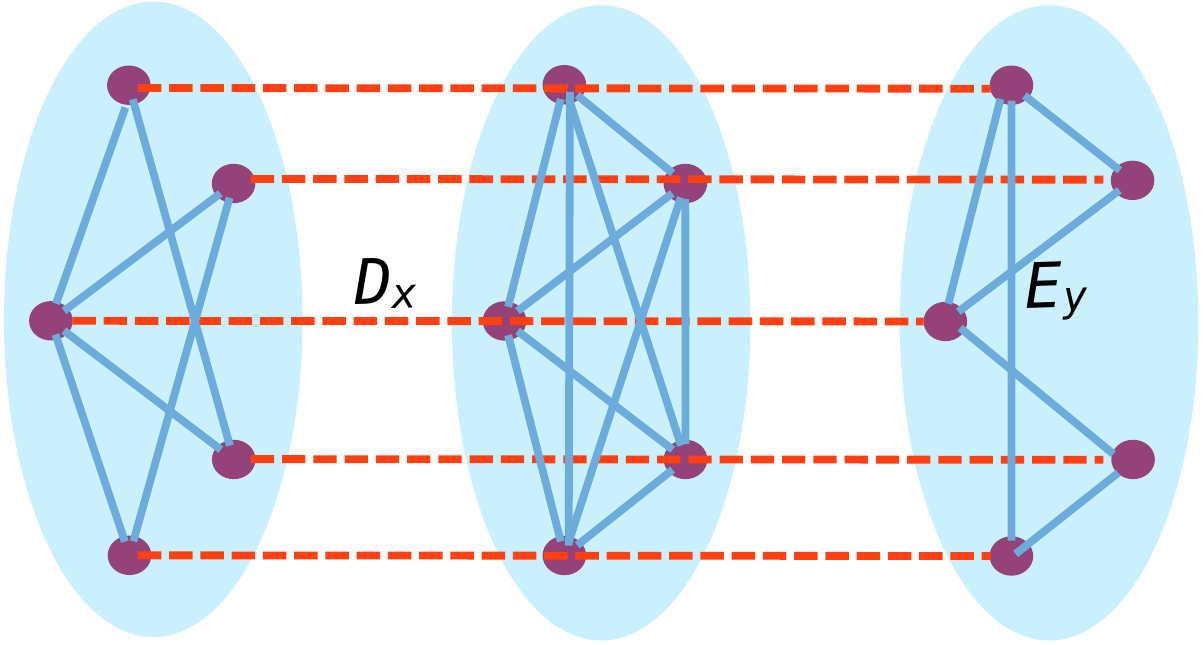}
		\caption{Schematic diagram of a 3-layered multiplex network comprised of a globally-connected and random networks. Blue solid lines denote intra-layer links while red hyphenated lines denote inter-layer links.}
	\label{mp_sch}
	\end{center}
\end{figure}
Next, we consider a multiplex network comprising of two layers of the same size $N$. Each node is represented by a second-order Kuramoto oscillators with its evolution governed by
\begin{align}
m\ddot\theta_{i,1} {+} \dot\theta_{i,1} = \omega_{i,1} {+} \frac{\lambda}{[\langle k_1\rangle{+}D_x]} \sum_{j{=}1}^{2N} M_{ij} \sin[\theta_{j,1}-\theta_{i,1}], \nonumber \\
m\ddot\theta_{i,2} {+} \dot\theta_{i,2} = \omega_{i,2} {+} \frac{\lambda}{[\langle k_{2}\rangle{+}D_x]} \sum_{{j=}1}^{2N} M_{ij} \sin[\theta_{j,2}-\theta_{i,2}],
\end{align}
where $i{=}1,...,N$, and subscripts $1, 2$ stand for the first and second layers, respectively. $D_x$ is inter-layer coupling strength. $M_{i,j}$ is element of adjacency matrix of multiplex network $M$ denoted as, 
\begin{equation}
M=\begin{bmatrix} A_1 & D_xI \\ D_xI & A_2E_y \end{bmatrix},
\end{equation}
where $E_y$ stands for intra-layer coupling strength of the second layer, $I$ is the identity matrix and $A_1$ and $A_2$ are adjacency matrices of the first and second layers, respectively.

\begin{figure}[t]
\begin{center}
\includegraphics[height=5cm,width=8.83cm]{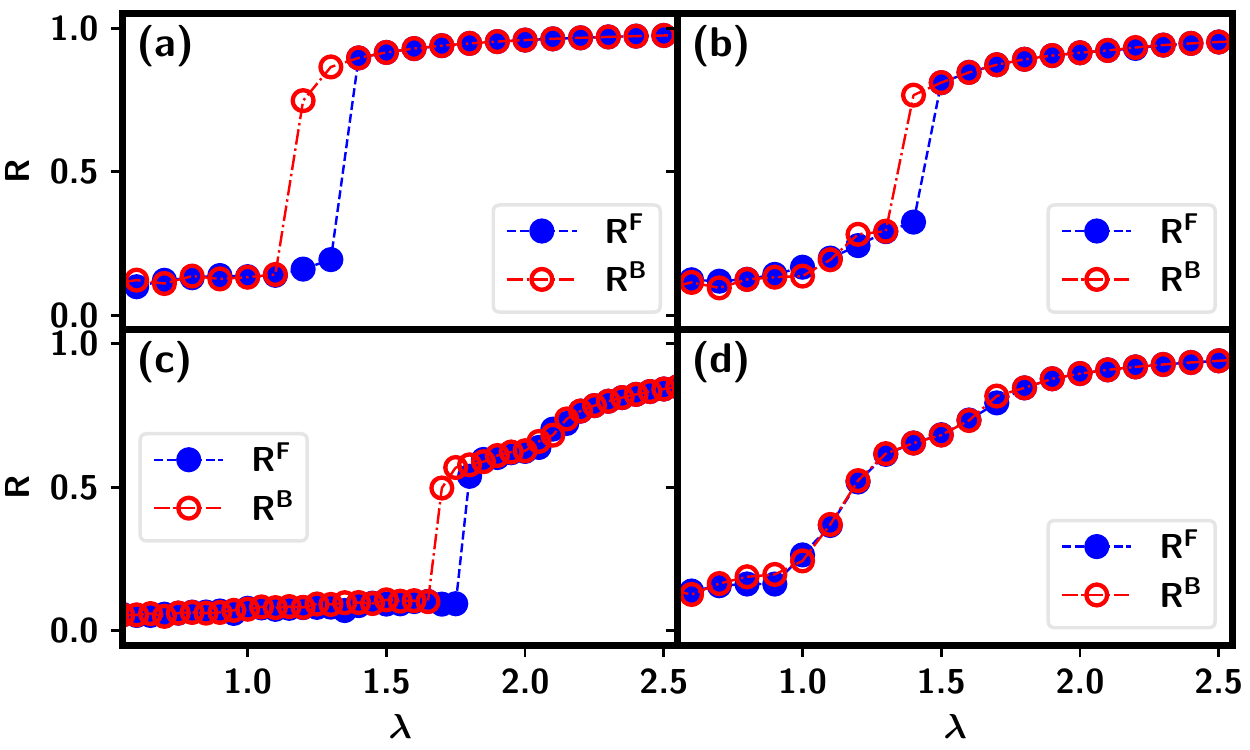}
\caption{Synchronization transition plots ($R$ vs. $\lambda$) of a single layer network for (a) globally connected, (b) ER, (c) WS and (d) SF network topologies. The size and average degree of each network is N=$100$ and $\langle k\rangle{=}12$.}
\label{fig:single_op}
\end{center}
\end{figure}
We track the phase transition to synchronization in both the layers of a multiplex network by means of order parameters $R_1$ and $R_2$, respectively, given by
\begin{align}
R_1e^{\iota\psi_1}=(1/N)\sum_{j=1}^{N}e^{\iota\theta_{j,1}},\nonumber \\
R_2e^{\iota\psi_2}=(1/N)\sum_{j=1}^{N}e^{\iota\theta_{j,2}}.
\end{align}

\begin{figure}[t]
	\begin{center}
		\includegraphics[height=4cm,width=5cm]{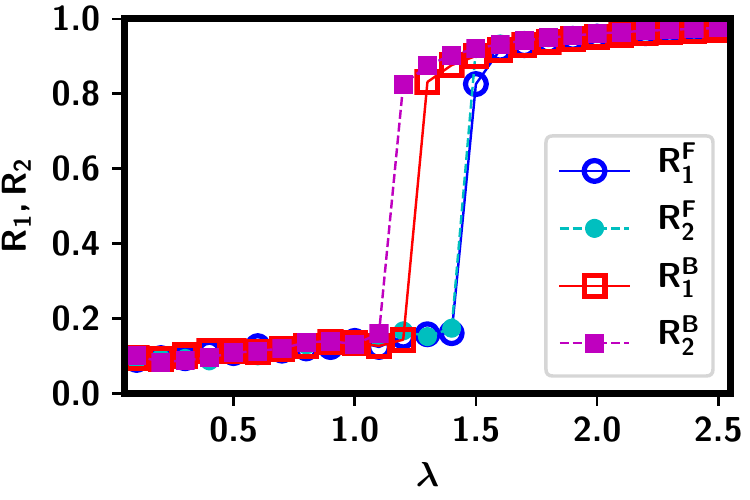}
		\caption{Synchronization plot of a multiplex network comprising two layers of globally connected network. The size of each layer is N=$100$.}
		\label{fig:mp_op_1gl2gl}
	\end{center}
\end{figure}
First, we adiabatically increase the coupling strength $\lambda$ from $\lambda_0$ (incoherent) to $\lambda_0{+}n\delta\lambda_0$ (synchronous) state in steps of $\delta\lambda_0$. Second, we adiabatically decrease the coupling strength from $\lambda_0{+}n\delta\lambda_0$ (synchronous) to $\lambda_0$ (incoherent) in the step of $\delta\lambda_0$. We compute the order parameters $R_1$ and $R_2$ for $\lambda_0,\lambda_0{+}\delta\lambda_0, ...\lambda_0{+}n\delta\lambda_0$ for both the increasing (forward, i.e.,\ $F$) and decreasing $\lambda$ (backward, i.e.,\ $B$) directions. Before each $\delta\lambda_0$ step, we eliminate initial transients and integrate the system long enough ($10^4$ time steps) using a fourth-order Runge-Kutta method with time step $dt{=}0.01$, to arrive at the stationary states. For all simulations, initial phases for oscillators in individual layer are selected from a random uniform distribution in the range $[0, 2\pi)$, and the natural frequencies of nodes of each network are drawn from a random uniform distribution from $[-1,1]$.

\section{Synchronization Transition in Monoplex Networks} 
Before presenting results for synchronization transition (ST) in a multiplex network, we discuss ST in a monoplex network having different network architectures such as globally-connected, Erd\"{o}s-R\'{e}nyi (ER) random, small-world (SW) and scale-free (SF) networks generated using Barab\'{a}si-Albert model \cite{netx}.\ Fig. \ref{fig:single_op} depicts behavior of order parameter $R$ as a function of the coupling strength for these different network models. It is evident that the globally connected, ER and SW networks exhibit a first-order transition with an associated hysteresis loop for the order parameters $R^F$ and $R^B$ indicating an occurrence of ES. Values of the critical couplings $\lambda_c^F$ and $\lambda_c^B$ in the forward and the backward directions, respectively, for the onset of ES are slightly different for the globally-connected, ER and SW networks, whereas the SF networks exhibit a usual second-order transition with critical coupling $\lambda_c{=}2.1$.

\section{Synchronization Transition in Multiplex Networks} Here we discuss investigation of ST in a multiplex network for various combinations of networks architecture for two layers. To systematically investigate the effects of multiplexing, we fix the first layer of the multiplex network to a globally connected network and choose architecture of the second layer from ER, SW and SF network models.

\begin{figure}[t]
	\begin{center}
		\includegraphics[height=3.5cm,width=8.83cm]{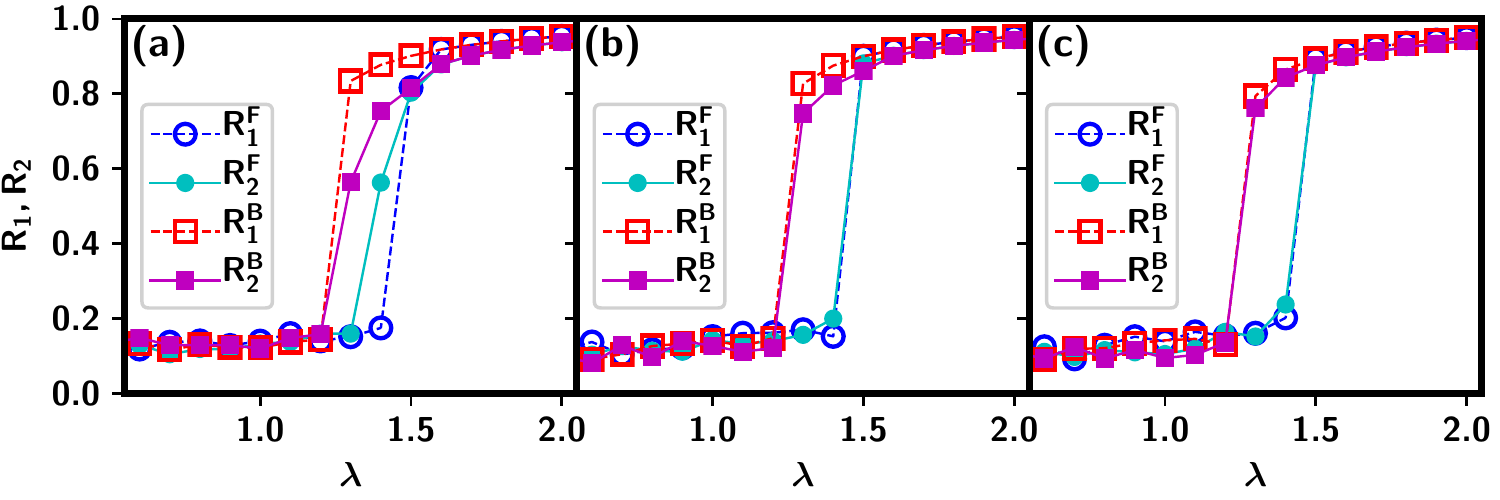}
		\caption{$R_1,R_2{-}\lambda$ plots of a multiplex network with the first layer fixed to globally connected network and the second layer to a ER network (a) $D_x{=}1$, (b) $D_x{=}4$ and (c) $D_x{=}8$. The size of each layer is N=$100$ and average degree of the second layer is $\langle k\rangle{=}12$.}
		\label{fig:mp_op_1gl2er}
	\end{center}
\end{figure}

\begin{figure}[t]
	\begin{center}
		\includegraphics[height=3.5cm,width=8.83cm]{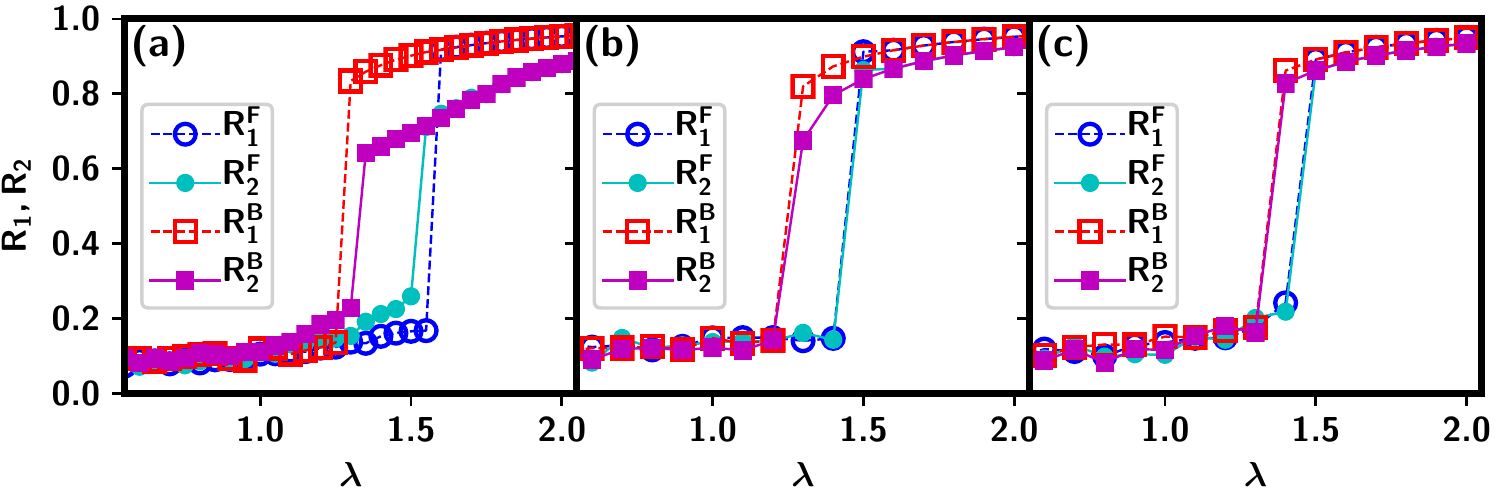}
		\caption{$R_1,R_2{-}\lambda$ plots of a multiplex network with the first layer fixed to globally connected network and the second layer to a SW  network (a) $D_x{=}1$, (b) $D_x{=}4$ and (c) $D_x{=}8$. The size of each layer is N=$100$ and average degree of the second layer is $\langle k\rangle{=}12$.}
		\label{fig:mp_op_1gl2ws}
	\end{center}
\end{figure}

\begin{figure}[htbp]
	\begin{center}
		\includegraphics[height=5cm,width=8.83cm]{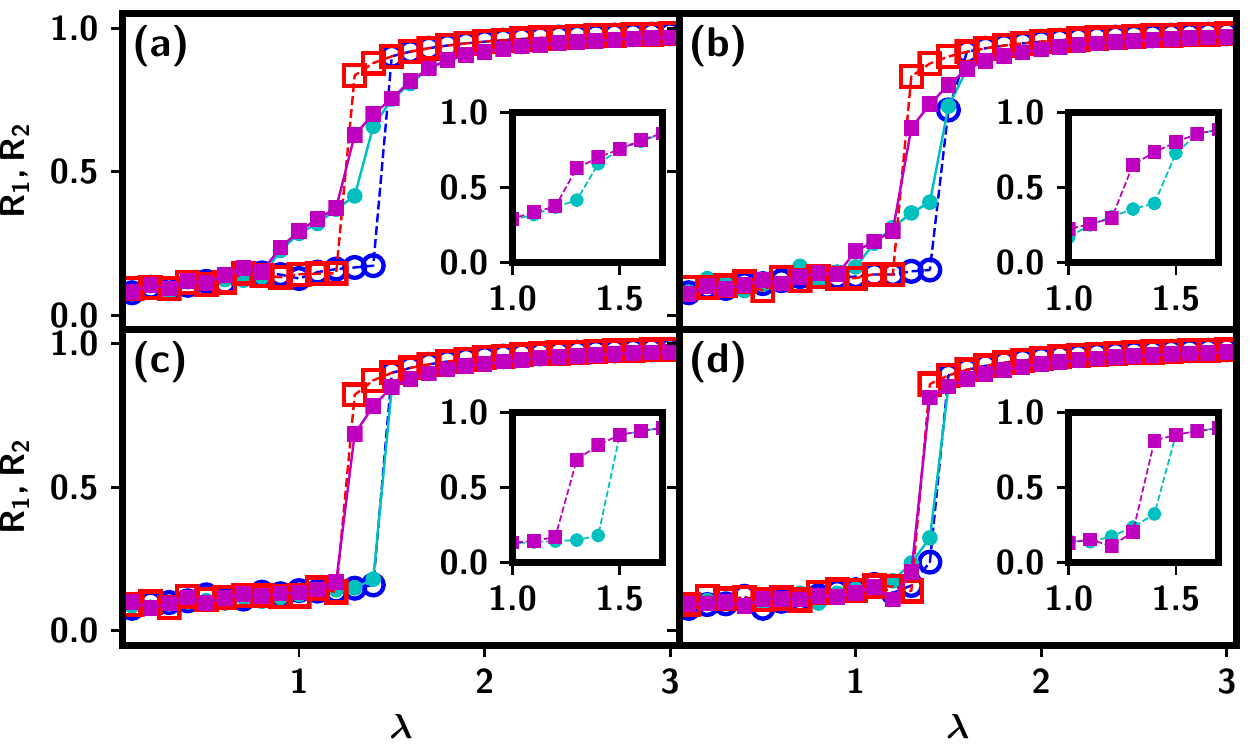}
		\caption{$R_1,R_2$ vs. $\lambda$ plots of a multiplex network with the first layer fixed to a globally connected network and the second layer to a SF network with inter-layer coupling strength (a) $D_x{=}1$, (b) $D_x{=}2$, (c) $D_x{=}4$ and (d) $D_x{=}8$. Symbols $\color{blue}\mbox{\Circpipe}$ and $\color{cyan}\mbox{\CIRCLE}$ denote $R_1$ and $R_2$ in the forward continuation while $\color{red}\mbox{\Squarepipe}$ and $\color{magenta}\blacksquare$ denote $R_1$ and $R_2$ in the backward continuation, respectively. The size of each layer is N=$100$ and average degree of the second layer is $\langle k\rangle{=}12$.}
		\label{fig:mp_op_1gl2netD}
	\end{center}
\end{figure}

For a multiplex network with the first and second layers represented by the globally connected networks, both the layers exhibit the first-order transitions and related hysteresis loops with values of $\lambda_c^F$ and $\lambda_c^B$ being $1.5$ and $1.3$, and $1.5$ and $1.2$, respectively (Fig. \ref{fig:mp_op_1gl2gl}). It indicates that upon multiplexing, the critical coupling $\lambda_c^F$ for both the layers increases and the hysteresis width either manifest an increase or remains the same. We further observe that when the second layer is comprised of either ER or SW  network, it exhibits the first-order (ES) transition and associated hysteresis loop for the forward and the backward directions in $\lambda$ (Fig.\ \ref{fig:mp_op_1gl2er}(a) and Fig.\ \ref{fig:mp_op_1gl2ws}(a)), respectively. Moreover, the hysteresis widths corresponding to the ER and SW  networks depict an enhancement upon multiplexing as compared to those for the corresponding monoplex networks (Fig. \ref{fig:single_op}(b) and Fig. \ref{fig:single_op}(c)).\ However, when the second layer is comprised of the SF network, multiplexing with a globally connected layer yields an interesting behavior. For this arrangement, the second layer also starts exhibiting a first-order (weak) transition with a hysteresis loop in the forward and in the backward direction (Fig. \ref{fig:mp_op_1gl2netD}(a)) as compared to the second-order transition manifested by the monoplex SF network (Fig. \ref{fig:single_op}(d)).

\noindent \textbf{Effect of inter-layer coupling strength on ST:} 
We have already witnessed that multiplexing two layers affects synchronization properties of both the layers. It is therefore interesting to trace that how variations in the inter-layer coupling strength further affect transition in the second layer.
First, we investigate the impact of inter-layer coupling strength on ST in a multiplex network with first layer fixed to a globally connected network and second layer represented by a ER network. For a usual case of inter-layer coupling strength being same as that of intra-layer (i.e.,\ $D_x{=}1$), ST in ER network exhibits the first-order transition with an existence of the hysteresis loop and the dynamical evolution of the ER layer does not synchronize simultaneously with that of the globally connected layer (see Fig. \ref{fig:mp_op_1gl2er}(a)). For a larger value of inter-layer coupling strength (say $D_x{=}4$), a stronger first-order transition is observed for the second layer. Additionally, the hysteresis loop for the second layer now exhibits a coherent behavior with that of the first layer (see Fig. \ref{fig:mp_op_1gl2er}(b)). For a further increase in inter-layer coupling strength the first-order ST for the second layer, corresponding to a large $D_x{=}8$, becomes as strong as that of the first layer and the hysteresis loop associated to it now is in a complete coherence with that of the first layer (see Fig. \ref{fig:mp_op_1gl2er}(c)).
Second, we investigate ST in the SW networks upon its multiplexing with a layer represented by globally connected network. It is found that for $D_x=1$, the SW layer displays the first-order transition (though weak as compared to the ER layer case) with a hysteresis loop which is not in coherence with that of the first layer as depicted in Fig. \ref{fig:mp_op_1gl2ws}(a). However, when inter-layer coupling $D_x$ is increased to $4$, the first-order ST observed for the SW layer becomes stronger. Moreover, both the layers now synchronize with each-other accompanied with a slight reduction in their hysteresis widths (Fig. \ref{fig:mp_op_1gl2ws}(b)). For a larger inter-layer coupling ($D_x{=}8$), the first-order ST for the second layer becomes almost as strong as that of the globally connected layer. Both the layers synchronize simultaneously and their hysteresis widths are further reduced to a new low as displayed in Fig. \ref{fig:mp_op_1gl2ws}(c). On the basis of above simulation results, we infer that strong inter-layer coupling strengthens the ES transition in a non-globally connected multiplex layer.

Next, we investigate effects of $D_x$ on ST in a multiplex network with the first layer fixed to a globally connected and the second layer to a SF network. As displayed by Fig. \ref{fig:mp_op_1gl2netD}(a), for a usual case of $D_x{=}1$, there is a weak first-order transition with a hysteresis loop in the forward and in the backward directions. Furthermore, for the SF layer, $R_2^F$ and $R_2^B$ do not get synchronized simultaneously with those of the globally connected first layer. Upon increasing $D_x$ to $2$, the first-order ST in the second layer gets enhanced or becomes stronger, and associated hysteresis width widens which gets synchronized almost simultaneously with the first layer (Fig. \ref{fig:mp_op_1gl2netD}(b)). For $D_x{=}4$, we observe an enhancement in ST of the second layer, and the second layer which now exhibits a stronger first-order transition accompanied with a broader hysteresis loop (Fig. \ref{fig:mp_op_1gl2netD}(c)). Moreover, it gets synchronized simultaneously with the first layer. For a further increase in $D_x$, a much stronger first-order ST is observed for the second layer displaying simultaneous synchronization with the first layer (Fig. \ref{fig:mp_op_1gl2netD}(d)). However, such a large inter-layer coupling ($D_x{=}8$) brings upon a reduction in the width of the synchronized hysteresis loops corresponding to both the first and the second layers. In this way, an increased $D_x$ with one layer having globally-connected architecture (yielding first-order ST) helps another layer having heterogeneous degree distribution (second-order ST) in achieving a first-order (explosive) ST. Therefore, choosing a high inter-layer coupling $D_x$ allows an enhancement in ST for multiplex networks.\\

\begin{figure}[t]
	\begin{center}
		\includegraphics[height=6cm,width=8.83cm]{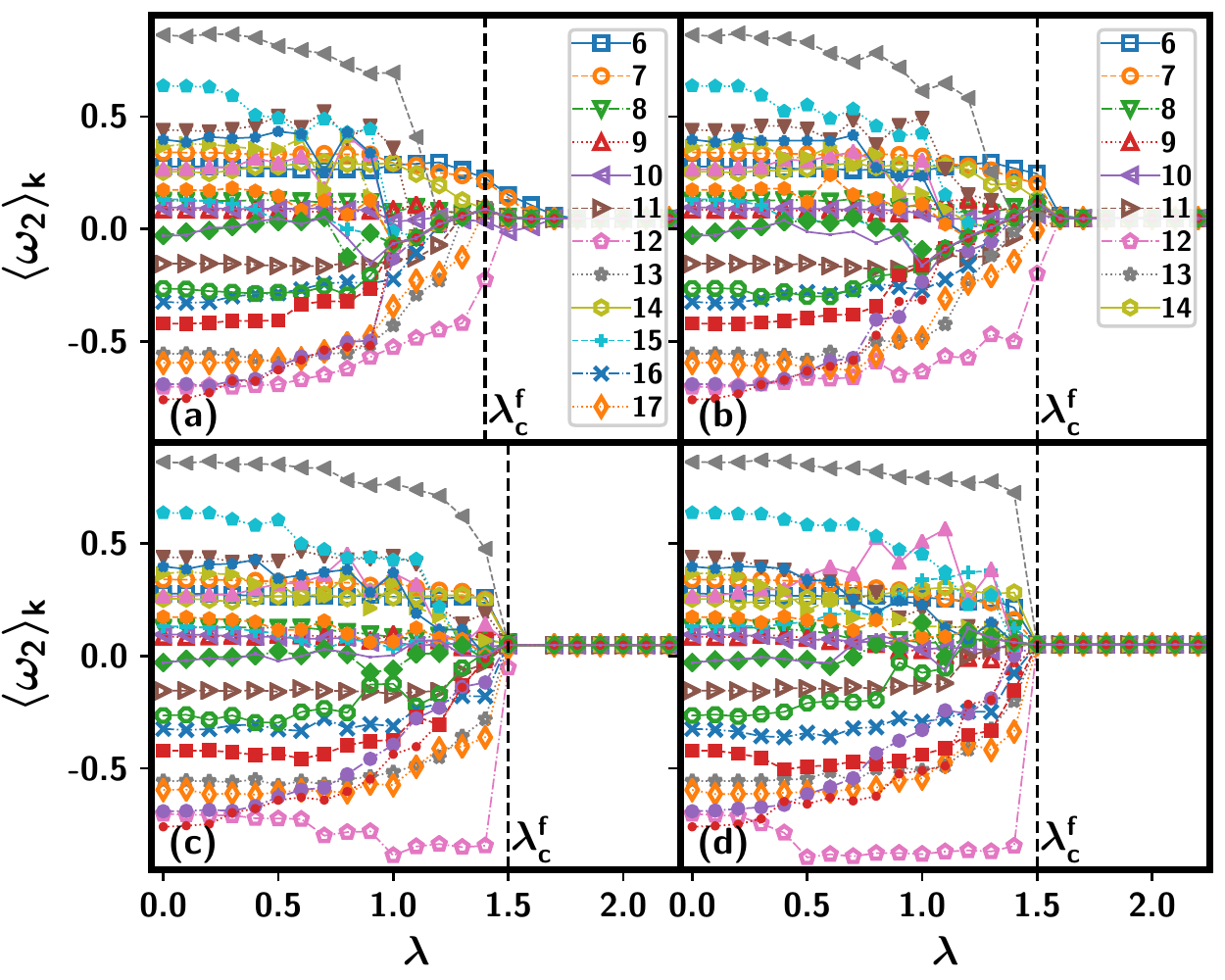}
		\caption{Average frequencies $\langle\omega_2\rangle_k$ of nodes in the second layer with the same degree $k$ as a function of $\lambda$ in a multiplex network with the first layer fixed to a globally connected network and the second layer to a SF network with inter-layer coupling (a) $D_x{=}1$, (b) $D_x{=}2$, (c) $D_x{=}4$ and (d) $D_x{=}8$. The numbers in plots (a) and (b) represent the degree of nodes of the SF network which get synchronized post critical-coupling $\lambda_c^f$. The size of each layer is N=$100$ and average degree of the second layer is $\langle k\rangle{=}12$.}
		\label{fig:mp_op_1gl2netMF}
	\end{center}
\end{figure}
To better understand the underlying dynamics behind the ES transition witnessed for the above case of a SF network multiplexed with a globally connected network, we perform detailed investigation of the nodes behavior. Since networks with the homogeneous and the heterogeneous degree-distributions manifest different behavior, we calculate average frequency of all the nodes with degree $k$. The average frequency for a degree $k$ is defined as :
\begin{equation}
\langle\omega\rangle_k=\sum_{i|k_i=k}\omega_i/N_k;\quad \omega_i=\int_t^{t+T}\dot\theta_i(t)dt/T,
\end{equation}
where $N_k$ is the number of nodes with degree $k$, and $T$ is the total time of averaging after eliminating initial transients. Fig. \ref{fig:mp_op_1gl2netMF} plots the average frequency $\langle\omega_2\rangle_k$ of all the nodes of degree $k$ in the SF network for different values of $D_x$. It is quite apparent that not all the nodes gets synchronized to the large synchronous component at the critical coupling $\lambda_c^f{=}1.4$. 
Fig. \ref{fig:mp_op_1gl2netMF}(a) infact indicates that the nodes with large degree converge first to a common average frequency at $\lambda_c^f$ while the nodes with small degree achieve synchronization post $\lambda_c^f$.\
A weak explosive synchronization is observed for $D_x{=}1$ as only a few nodes are synchronized subsequent to the critical coupling $\lambda_c^f$. The number of nodes synchronizing post the critical coupling $\lambda_c^f{=}1.5$ gets further reduced for $D_x{=}2$ (Fig. \ref{fig:mp_op_1gl2netMF}(b)). For the stronger couplings $D_x{=}4$ and $D_x{=}8$, nodes of all the degrees synchronize simultaneously at the critical coupling $\lambda_c^f{=}1.5$ (as depicted in Fig. \ref{fig:mp_op_1gl2netMF}(c) and Fig. \ref{fig:mp_op_1gl2netMF}(d)), yielding an obvious first-order (explosive) transition.\ 
It is apparent that the larger $D_x$ values do not influence the onset of synchronization as the value of $\lambda_c^F$ remains fixed to $1.5$ for higher values of $D_x$ in a network of 100 nodes.\ From these observations we can conclude that strengthening the inter-layer coupling (or multiplexing impact) greatly enhances the underlying dynamics towards ES.\\

\begin{figure}[t]
	\begin{center}
		\includegraphics[height=6cm,width=8.83cm]{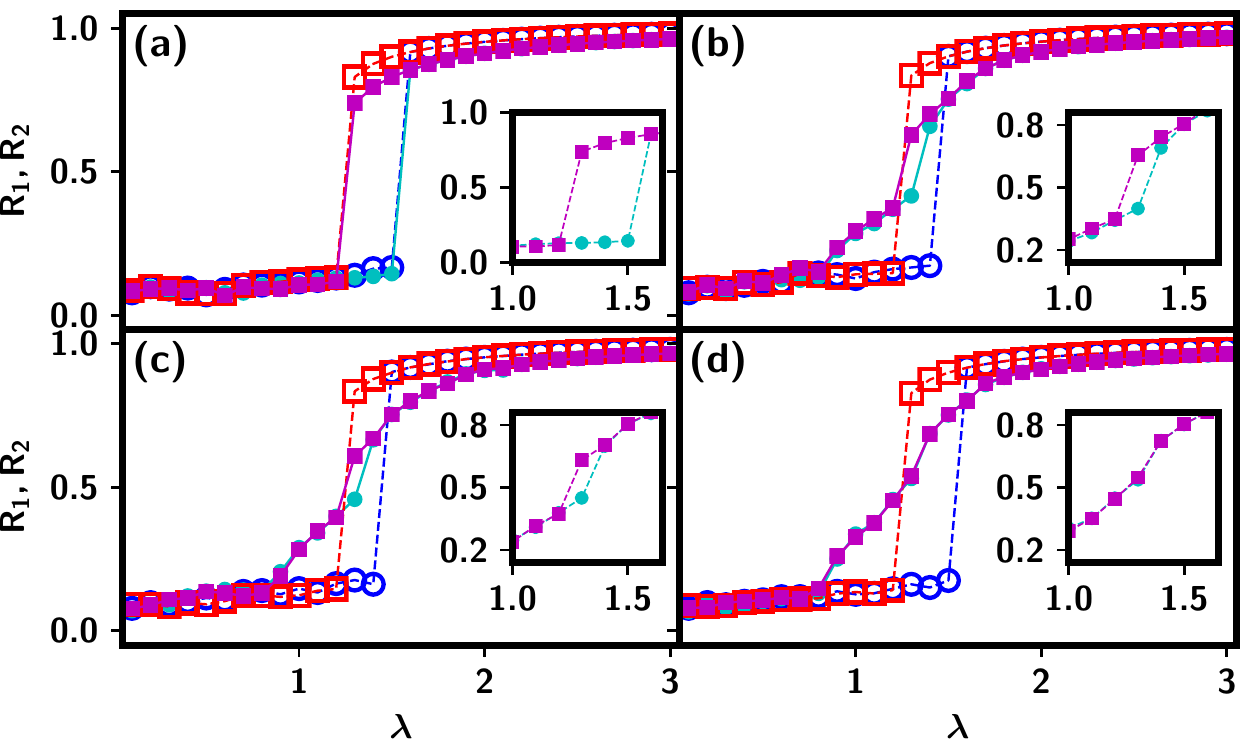}
		\caption{$R_1, R_2$ vs. $\lambda$ plots of a multiplex network with the first layer fixed to a globally connected network and the second layer to a SF network with $D_x{=}1$ and (a) $E_y{=}0$, (b) $E_y{=}1$, (c) $E_y{=}4$ and (d) $E_y{=}8$. Symbols $\color{blue}\mbox{\Circpipe}$ and $\color{cyan}\mbox{\CIRCLE}$ denote $R_1$ and $R_2$ in the forward direction, while $\color{red}\mbox{\Squarepipe}$ and $\color{magenta}\blacksquare$ denote $R_1$ and $R_2$ in the backward direction, respectively. The size of each layer is N=$100$ and average degree of the second layer is $\langle k\rangle{=}12$.}
		\label{fig:mp_op_1gl2netE}
	\end{center}
\end{figure}

\noindent \textbf{Effect of intra-layer coupling strength on ST:} Next, we discuss impact of relative mis-match in intra-layer coupling for layer, on ST for multiplex networks. Let us first consider a multiplex network comprising a globally connected and a SF networks. $E_y$ is introduced in the SF layer of the multiplex network. As we have already seen that for a usual choice of $D_x{=}1$ and $E_y{=}1$ (Fig. \ref{fig:mp_op_1gl2netE}(b)), the first layer having globally connected and the second layer having SF topology display a strong discontinuous and weakly discontinuous transitions 
(seen at $\lambda_c^F{=}1.4$), respectively. However, when $E_y$ is set to $0$, which corresponds to each node of the second layer being connected mere to a single node in the first layer via the inter-layer couplings, this arrangement leads to an enhancement in ST in the second layer causing a strong first-order transition (see Fig. \ref{fig:mp_op_1gl2netE}(a)) which synchronizes simultaneously with the first layer. On contrary, for an increased $E_y$ (say $E_y{=}4$), a weak discontinuous ST in the SF network becomes further weaker with no change in the onset of synchronization or $\lambda_c^F ({=}1.4)$ (see Fig. \ref{fig:mp_op_1gl2netE}(c)). For a more strong $E_y$, the weak first-order ST with its hysteresis loop gets disappeared and instead a second order transition is observed. Therefore, a gradual increase in the intra-layer coupling strength weakens the first-order (explosive) ST further and eventually turns it into a second-order one with no hysteresis. Hence, a lower value of $E_y$ in heterogeneous layer is good for the enhancement of explosive synchronization in the same layer.

\begin{figure}[t]
	\begin{center}
		\includegraphics[height=4cm,width=8.83cm]{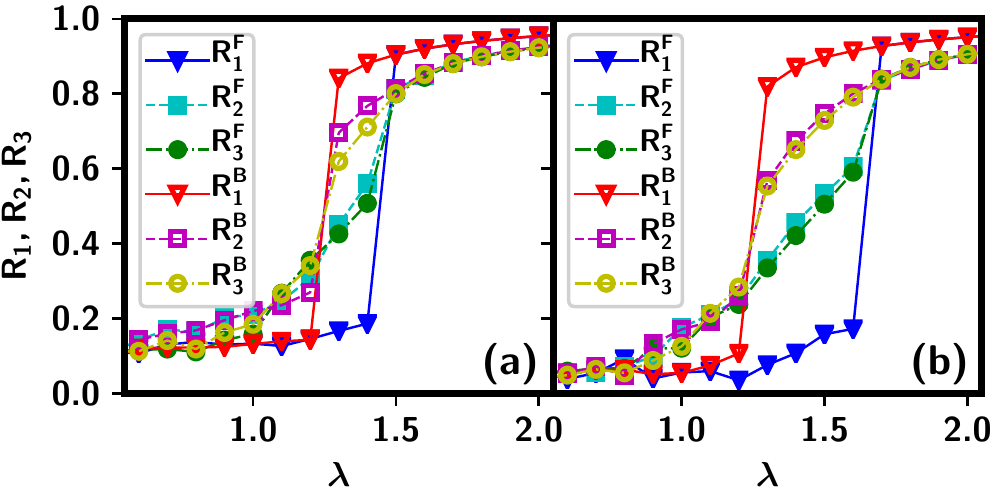}
		\caption{$R_1,R_2,R_3{-}\lambda$ plots of a 3-layered multiplex network comprised of a globally connected and two scale-free layers, each of (a) $100$ and (b) $500$ nodes. The average degrees of the second and third layers are $\langle k\rangle{=}12$.}
		\label{fig:mp_op_glbaba_3l}
	\end{center}
\end{figure}

\noindent \textbf{{Effect of number of layers and size of network on ST:}} It is important to investigate how the nature of synchronization is affected by change in both the number of layers and the number of nodes in multiplex networks. 
Fig.\ \ref{fig:mp_op_glbaba_3l} presents ST for 3-layered multiplex networks corresponding to two different network size of $100$ and $500$ nodes. We present results for multiplex network consisting of one globally connected and two SF layers.\ It transpires that the globally connected layer sticks to its ES nature while the two SF layers exhibit weak ES transition ($R_2$ and $R_3$).\ Therefore, the nature of synchronization for a 3-layered multiplex network is similar to that observed for the case of 2-layered network comprising of the globally connected and SF layers.\ However, interestingly, a single layer of globally connected network (having homogeneous degree-distribution) is capable of inducing ES transition in both the SF layers (having heterogeneous degree-distribution).\ Besides, the size of network has no significant effect on the nature of ST and only affects the onset of synchronization, i.e., the critical coupling strength.\ For instance, Fig.\ \ref{fig:mp_op_glbaba_3l} indicates that the pair of values of the critical coupling strengths \{$\lambda^F$, $\lambda^B$\} for network size of $100$ and $500$ are \{$1.5, 1.3$\} and \{$1.7, 1.3$\}, respectively.

\begin{figure}[t]
	\begin{center}
		\includegraphics[height=3.5cm,width=8.83cm]{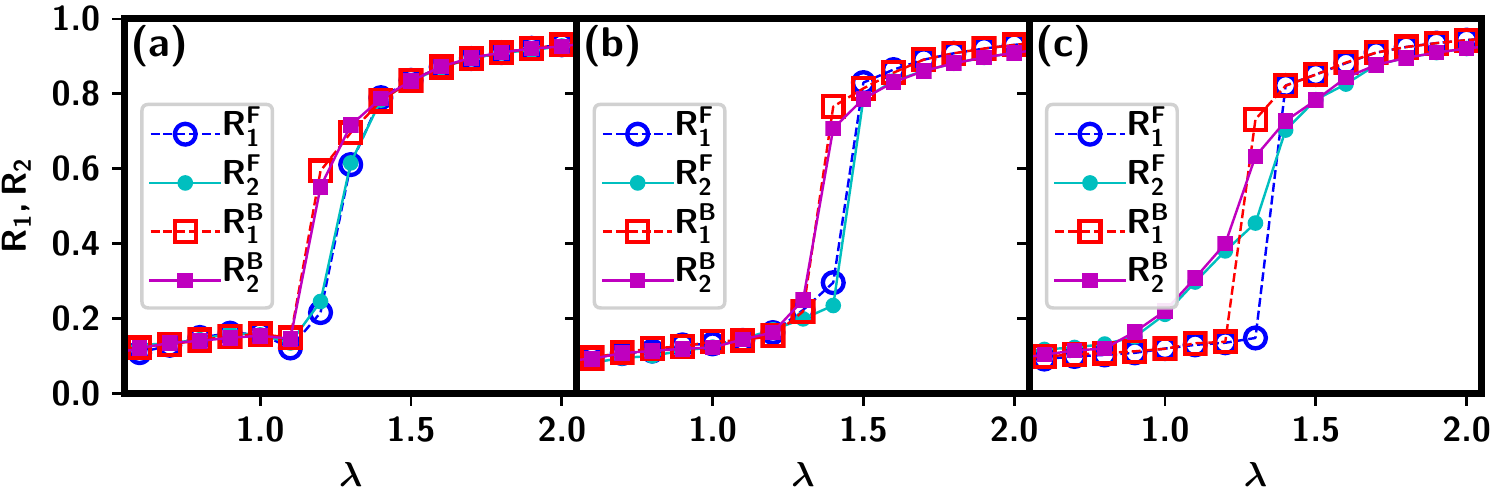}
		\caption{Synchronization diagrams of a multiplex network with the first layer fixed to ER network and the second layer to a (a) ER (b) SW  and (c) BA network. The size and average degree of each layer's network is N=$100$ and $\langle k\rangle{=}12$.}
		\label{fig:mp_op_netx}
	\end{center}
\end{figure}
\noindent \textbf{Effect of non-globally connected layers on ST:} Next, we consider both layers of a multiplex network represented by non-globally connected networks. We consider a multiplex network consisting of the first layer fixed to a ER network and the second layer is chosen from ER, SW or a SF network. However for these case, behavior of ST is found to be similar to that of the respective cases of multiplex networks with first layer fixed to a globally connected network. Since, both the globally connected and ER monoplex networks bear homogeneous degree-distributions and display a first-order (explosive) ST. From Fig. \ref{fig:mp_op_netx}(a) and Fig. \ref{fig:mp_op_netx}(b), it is quite apparent that the second layer comprising ER or SW  network gets synchronized simultaneously with the first layer, with the forward and the backward transitions in $\lambda$ exhibiting hysteresis loops. However, when the second layer is comprised of a SF network, it exhibits a first-order (weak) transition with associated hysteresis loop in the forward and the backward directions (Fig.\ \ref{fig:mp_op_netx}(c)). This is an interesting revelation as this combination of layers multiplexed with each-other displays ST behavior similar to that seen in the case of a multiplex network comprising the globally connected and the SF layers. Hence, we deduce that multiplexing a layer having heterogeneous degree-distribution to an another layer having homogeneous degree-distribution induces the first-order (ES) transition in the former layer which was displaying a second-order transition in isolation. It remains to be investigated further that why a first-order transition always gets induced in the heterogeneous layer and the reverse, i.e.,\ a second-order transition induced in the homogeneous layer, does not occur.

\section{Discussions}
In this article, we explored synchronization transition in second-order Kuramoto oscillators on multiplex network and focused to understand how network structure in one layer affects the synchronization transition in another layer. We observe that the second-order Kuramoto oscillators on monoplex network having homogeneous degree-distribution such as globally connected, ER or SW network displays first order (ES) transition, while the coupled dynamics on networks having heterogeneous degree-distribution, such as SF networks displays a second-order phase transition. We have reported that when a layer with homogeneous degree-distribution is multiplexed with a second layer bearing a similar type of degree-distribution, both layers display a first-order ST. However, a network which follows a second-order transition in isolation such as the networks which are highly heterogeneous, upon multiplexing with another network having homogeneous degree-distribution, starts displaying a weak first-oder ST. Hence, the first-order transition in the second-order Kuramoto oscillators can be induced via multiplexing in those networks which exhibit a second-order ST in isolation. Therefore, in a two-layered multiplex network if one layer comprises a network with a homogeneous degree-distribution another layer will display a first-order transition irrespective of its distribution type. It is further numerically demonstrated  that such induced rather weak ES transition can be enhanced making it strong by controlling the inter-layer coupling. Similarly, it can be weakened further eventually turning into a second-order transition by controlling relative intra-layer couplings of layers of the multiplex network. 

It has been sought to reason the occurrence of ES. For instance, it has been shown that coexistence of disassortativity in the degrees and the natural frequencies of the nodes leads to ES \cite{zhu}. Furthermore, it was demonstrated that the dynamical origin of the hysteresis corresponding to ES in a heterogeneous network results from the change in the basin of attraction of the synchronized state \cite{zou}.\ Moreover, using Ott-Antonsen approach \cite{ott}, it was reported that in low-dimensional dynamical space corresponding to the globally synchronized state in heterogeneous networks undergo various transitions among diverse collective states such as fixed points and limit cycles, and ES is determined by the bistable state \cite{canxu}.
The reason for witnessing the first- and second-order transitions for the SF layer in the 
multiplex network considered here lies in the heterogeneity of degrees, which we have explained by computing the average frequencies of nodes having the same degree in the SF layer.\ When a SF layer is multiplexed with a globally connected layer, the hubs or higher degree nodes in the SF layer get synchronized first. However, the inter-layer coupling strength ($D_x{=}1$) is not strong enough for the lower degree nodes, driven by globally connected nodes, to get synchronized.\ Therefore, we observe a weak ES transition.\ When $D_x$ is sufficiently high (say $D_x{=}4$), the lower degree nodes also, along with the hubs, driven by the globally connected nodes, get synchronized leading to a strong ES transition.

To conclude, earlier studies have shown that coupled dynamical behaviors such as cluster 
synchronization of a layer can be governed by network properties of other layers in multiplex network 
\cite{aradhana}. Here, we demonstrate that ES of one or more than one layer can be achieved by 
multiplexing them with an appropriate network structure.

The Kuramoto oscillators with inertia provide a more realistic model of many systems such power-grid \cite{fila,dorf,pinto}, disordered arrays of Josephson junctions \cite{trees}, etc.\ In a power grid, the first- and second-order Kuramoto oscillators correspond to loads and generators, respectively. These realization have lead a spurt in the activities of investigation of second-order Kuramoto oscillators on networks to get insights to underlying mechanism behind emerging phenomena due to interactions of nonlinear dynamical units. All these investigations on second-order Kuramoto oscillators are restricted to monoplex networks. This article investigates impact of multiplexing on behavior of second-order Kuramoto oscillators, and shows that there is a relation between the strength of multiplexing (inter-layer coupling strength) and ST. The results are important to understand such transitions observed in real-world complex systems inherently having multi-layers architecture.

\acknowledgments SJ acknowledges DST, Government of India project grant EMR/2016/001921 for financial support.



\begin{thebibliography}{0}
\bibitem{netx}{In Erd\"{o}s-R\'{e}nyi (ER) random network, nodes are connected by edges chosen at random with probability $0{<}P{<}1$ \cite{erdos}. A Watts-Strogatz (SW ) small-world network is generated by rewiring edges at random with probability $0{<}p{<}1$, in a ring lattice with nearest neighbors connections \cite{watts}. In the scale-free (BA model) networks, new nodes are added in accordance with preferential attachment and are characterized by a power-law degree-distribution \cite{barabasi}.}
\end{thebibliography}
\end{document}